\documentclass[acus]{JAC2001}

\usepackage{graphicx}

\begin{document}

\title{HAMILTONIAN FORMALISM FOR SOLVING THE VLASOV-POISSON 
EQUATIONS AND ITS APPLICATION TO THE COHERENT BEAM-BEAM 
INTERACTION} 

\author{Stephan I. Tzenov and Ronald C. Davidson\\ 
Plasma Physics Laboratory, Princeton University, Princeton, 
New Jersey 08543, USA}

\maketitle

\*\vspace{0.2cm}

\begin{abstract}
A Hamiltonian approach to the solution of the Vlasov-Poisson equations 
has been developed. Based on a nonlinear canonical transformation, the 
rapidly oscillating terms in the original Hamiltonian are transformed 
away, yielding a new Hamiltonian that contains slowly varying terms 
only. The formalism has been applied to the coherent beam-beam 
interaction, and a stationary solution to the transformed Vlasov 
equation has been obtained. 
\end{abstract}

\renewcommand{\theequation}{\thesection.\arabic{equation}}

\setcounter{equation}{0}

\*\vspace{0.0cm}

\section{INTRODUCTION}

The evolution of charged particle beams in accelerators and storage 
rings can often be described by the Vlasov-Maxwell equations. At high 
energies the discrete-particle collision term \cite{tzenov1} comprises 
a small correction to the dynamics and can be neglected. Radiation 
effects at sufficiently high energies for leptons can be a significant 
feature of the dynamics, and should be included in the model under 
consideration.

The Vlasov-Maxwell equations constitute a considerable simplification 
in the description of charged particle beam propagation. Nonetheless 
there are only a few cases that are tractable analytically. Therefore, 
it is of utmost the importance to develop a systematic perturbation 
approach, able to provide satisfactory results in a wide variety of 
cases of physical interest. 

Particle beams are subject to external forces that are often rapidly 
oscillating, such as quadrupolar focusing forces, RF fields, etc. In 
addition, the collective self-field excitations can be rapidly 
oscillating as well. A typical example is a colliding-beam storage 
ring device, where the evolution of each beam is strongly affected 
by the electromagnetic force produced by the counter-propagating beam. 
The beam-beam kick each beam experiences is localized only in a small 
region around the interaction point, and is periodic with a period of 
one turn. 

In this and other important cases one is primarily interested in the 
long-time behavior of the beam, thus discarding the fast processes on 
time scales of order the period of the rapid oscillations. To extract 
the relevant information, an efficient method of averaging is developed 
in the next section. Unlike the standard canonical perturbation 
technique \cite{channell,qin}, the approach used here is carried out 
in a ``mixed'' phase space (old coordinates and new canonical momenta), 
which is simpler and more efficient in a computational sense. The 
canonical perturbation method devel- 

\*\vspace{1.0cm} 

\noindent 
oped here is further applied to the 
coherent beam-beam interaction, and a coupled set of nonlinear integral 
equations for the equilibrium beam densities has been derived. 

\renewcommand{\theequation}{\thesection.\arabic{equation}}

\setcounter{equation}{0}

\section{THE HAMILTONIAN FORMALISM} 

We consider a $N$-dimensional dynamical system, described by the 
canonical conjugate pair of vector variables 
${\left( {\bf q}, {\bf p} \right)}$ with components 
\begin{eqnarray}\label{eq:components} 
{\bf q} = {\left( q_1, q_2, \dots, q_N \right)}, 
\nonumber \\ 
{\bf p} = {\left( p_1, p_2, \dots, p_N \right)}. 
\end{eqnarray} 
\noindent 
The Vlasov equation for the distribution function 
$f{\left( {\bf q}, {\bf p}; t \right)}$ can be expressed as 
\begin{equation}\label{eq:vlasov} 
{\frac {\partial f} {\partial t}} + 
{\left[ f, H \right]} = 0, 
\end{equation} 
\noindent 
where 
\begin{equation}\label{eq:bracket} 
{\left[ F, G \right]} = 
{\frac {\partial F} {\partial q_i}} 
{\frac {\partial G} {\partial p_i}} - 
{\frac {\partial F} {\partial p_i}} 
{\frac {\partial G} {\partial q_i}} 
\end{equation} 
\noindent 
is the Poisson bracket, $H{\left( {\bf q}, {\bf p}; t \right)}$ is 
the Hamiltonian of the system, and summation over repeated indices is 
implied. Next we define a canonical transformation via the generating 
function of the second type according to 
\begin{equation}\label{eq:canonical} 
S = S {\left( {\bf q}, {\bf P}; t \right)}, 
\end{equation} 
\noindent 
and assume that the Jacobian matrix 
\begin{equation}\label{eq:jacobian} 
{\cal J}_{ij} {\left( {\bf q}, {\bf P}; t \right)} = 
{\frac {\partial^2 S} {\partial q_i \partial P_j}} 
\end{equation} 
\noindent 
is non-degenerate with 
\begin{equation}\label{eq:degenerate} 
\det {\left( {\cal J}_{ij} \right)} \neq 0, 
\end{equation} 
\noindent 
so that the inverse ${\cal J}_{ij}^{-1}$ exists. Let us also define 
the distribution function in terms of the new coordinates 
${\left( {\bf Q}, {\bf P} \right)}$ and the mixed pair 
${\left( {\bf q}, {\bf P} \right)}$ as
\begin{equation}\label{eq:distrib} 
f{\left( {\bf q}, {\bf p}; t \right)} = 
f_0{\left( {\bf Q}, {\bf P}; t \right)} = 
F_0{\left( {\bf q}, {\bf P}; t \right)}. 
\end{equation} 
\noindent 
The new canonical variables ${\left( {\bf Q}, {\bf P} \right)}$ are 
defined by the canonical transformation as 
\begin{equation}\label{eq:transform} 
p_i = {\frac {\partial S} {\partial q_i}}, 
\qquad \qquad 
Q_i = {\frac {\partial S} {\partial P_i}}. 
\end{equation} 
\noindent 
Because 
\begin{equation}\label{eq:intermed1} 
{\frac {\partial p_i} {\partial P_j}} = 
{\frac {\partial^2 S} {\partial q_i \partial P_j}} = 
{\cal J}_{ij} \qquad \Longrightarrow \qquad 
{\frac {\partial P_i} {\partial p_j}} = 
{\cal J}_{ij}^{-1}, 
\end{equation} 
\noindent 
we can express the Poisson bracket in terms of the mixed variables 
in the form 
\begin{equation}\label{eq:bracketmix} 
{\left[ F, G \right]} = 
{\cal J}_{ji}^{-1} {\left( 
{\frac {\partial F} {\partial q_i}} 
{\frac {\partial G} {\partial P_j}} - 
{\frac {\partial F} {\partial P_j}} 
{\frac {\partial G} {\partial q_i}} 
\right)}. 
\end{equation} 
\noindent 
Differentiation of Eq. (\ref{eq:transform}) with respect to time 
$t$, keeping the old variables ${\left( {\bf q}, {\bf p} \right)}$ 
fixed, yields 
\begin{equation}\label{eq:intermed2} 
{\frac {\partial^2 S} {\partial q_i \partial t}} + 
{\frac {\partial^2 S} {\partial q_i \partial P_j}} 
{\left( {\frac {\partial P_j} {\partial t}} 
\right)}_{qp} = 0, 
\end{equation} 
\begin{equation}\label{eq:intermed3} 
{\left( {\frac {\partial Q_i} {\partial t}} 
\right)}_{qp} = 
{\frac {\partial^2 S} {\partial P_i \partial t}} + 
{\frac {\partial^2 S} {\partial P_i \partial P_j}} 
{\left( {\frac {\partial P_j} {\partial t}} 
\right)}_{qp}, 
\end{equation} 
\noindent 
or 
\begin{equation}\label{eq:intermed4} 
{\left( {\frac {\partial P_j} {\partial t}} 
\right)}_{qp} = - {\cal J}_{ji}^{-1} 
{\frac {\partial^2 S} {\partial q_i \partial t}}. 
\end{equation} 
\noindent 
Our goal is to express the Vlasov equation (\ref{eq:vlasov}) in terms 
of the mixed variables ${\left( {\bf q}, {\bf P} \right)}$. Taking 
into account the identities 
\begin{equation}\label{eq:intermed5} 
{\frac {\partial Q_i} {\partial q_j}} = 
{\frac {\partial^2 S} {\partial q_j \partial P_i}} = 
{\cal J}_{ji} \quad \Longrightarrow \quad 
{\frac {\partial q_i} {\partial Q_j}} = 
{\cal J}_{ji}^{-1}, 
\end{equation} 
\begin{equation}\label{eq:intermed6} 
{\frac {\partial f_0} {\partial Q_i}} = 
{\cal J}_{ij}^{-1} 
{\frac {\partial F_0} {\partial q_j}}, 
\end{equation} 
\begin{equation}\label{eq:intermed7} 
{\frac {\partial f_0} {\partial P_i}} = 
{\frac {\partial F_0} {\partial P_i}} - 
{\frac {\partial f_0} {\partial Q_j}} 
{\frac {\partial^2 S} {\partial P_i \partial P_j}}, 
\end{equation} 
\noindent 
we obtain 
\begin{eqnarray} 
{\left( {\frac {\partial f} {\partial t}} 
\right)}_{qp} = 
{\frac {\partial f_0} {\partial t}} + 
{\frac {\partial f_0} {\partial Q_i}} 
{\left( {\frac {\partial Q_i} {\partial t}} 
\right)}_{qp} + 
{\frac {\partial f_0} {\partial P_i}} 
{\left( {\frac {\partial P_i} {\partial t}} 
\right)}_{qp} \nonumber 
\end{eqnarray} 
\begin{eqnarray} 
= {\frac {\partial F_0} {\partial t}} + 
{\cal J}_{ji}^{-1} {\left( 
{\frac {\partial F_0} {\partial q_i}} 
{\frac {\partial^2 S} {\partial t \partial P_j}} - 
{\frac {\partial F_0} {\partial P_j}} 
{\frac {\partial^2 S} {\partial t \partial q_i}} 
\right)} \nonumber 
\end{eqnarray} 
\begin{equation}\label{eq:vlasovmix} 
= {\frac {\partial F_0} {\partial t}} + 
{\left[ F_0, {\frac {\partial S} {\partial t}} 
\right]}. 
\end{equation} 
\noindent 
Furthermore, using the relation 
\begin{equation}\label{eq:relation} 
{\left[ f, H \right]} = 
{\left[ F_0, {\cal H} \right]}, 
\end{equation} 
\noindent 
where 
\begin{equation}\label{eq:newhamil} 
{\cal H} {\left( {\bf q}, {\bf P}; t \right)} = 
H {\left( {\bf q}, {\bf \nabla}_q S; t \right)}, 
\end{equation} 
\noindent 
we express the Vlasov equation in terms of the mixed variables 
according to 
\begin{equation}\label{eq:mixvlasov} 
{\frac {\partial F_0} {\partial t}} + 
{\left[ F_0, {\cal K} \right]} = 0, 
\end{equation} 
\noindent 
where
\begin{equation}\label{eq:newhamilt} 
{\cal K} {\left( {\bf q}, {\bf P}; t \right)} = 
{\frac {\partial S} {\partial t}} + 
H {\left( {\bf q}, {\bf \nabla}_q S; t \right)} 
\end{equation} 
\noindent 
is the new Hamiltonian. For the distribution function 
$f_0 {\left( {\bf Q}, {\bf P}; t \right)}$ depending on the new 
canonical variables, we clearly obtain 
\begin{equation}\label{eq:equilib} 
{\frac {\partial f_0} {\partial t}} + 
{\left[ f_0, {\cal K} \right]} = 0, 
\end{equation} 
\noindent 
where the new Hamiltonian ${\cal K}$ is a function of the new 
canonical pair ${\left( {\bf Q}, {\bf P} \right)}$, such that 
\begin{equation}\label{eq:newhamilto} 
{\cal K} {\left( {\bf \nabla}_P S, {\bf P}; t \right)} = 
{\frac {\partial S} {\partial t}} + 
H {\left( {\bf q}, {\bf \nabla}_q S; t \right)}, 
\end{equation} 
\noindent 
and the Poisson bracket entering Eq. (\ref{eq:equilib}) has the same 
form as Eq. (\ref{eq:bracket}), expressed in the new canonical 
variables. 

\renewcommand{\theequation}{\thesection.\arabic{equation}}

\setcounter{equation}{0}

\section{COHERENT BEAM-BEAM INTERACTION} 

As an application of the formalism developed in the previous section 
we study here the evolution of two counter-propagating beams, 
nonlinearly coupled by the electromagnetic interaction between the 
beams at collision. For simplicity, we consider one-dimensional 
motion in the vertical $(q)$ direction, described by the nonlinear 
Vlasov-Poisson equations 
\begin{equation}\label{eq:vlasovk} 
{\frac {\partial f_k} {\partial \theta}} + 
{\left[ f_k, H_k \right]} = 0, 
\end{equation} 
\begin{equation}\label{eq:poisson} 
{\frac {\partial^2 V_k} {\partial q^2}} = 
4 \pi \! \! \int \! \! dp f_{3-k} (q, p; \theta), 
\end{equation} 
\noindent 
where 
\begin{equation}\label{eq:hamiltonian} 
H_k = {\frac {\nu_k} {2}} 
{\left( p^2 + q^2 \right)} + 
\lambda_k \delta_p (\theta) 
V_k (q; \theta) 
\end{equation} 
\noindent 
is the Hamiltonian. The notation in Eqs. (\ref{eq:vlasovk}) - 
(\ref{eq:hamiltonian}) is the same as in Ref. \cite{tzenov}. Our 
goal is to determine a canonical transformation such that the new 
Hamiltonian is time-independent. As a consequence, the stationary 
solution of the Vlasov equation (\ref{eq:mixvlasov}) is expressed 
as a function of the new Hamiltonian. Following the procedure 
outlined in the preceding section we transform Eqs. 
(\ref{eq:vlasovk}) - (\ref{eq:hamiltonian}) according to 
\begin{equation}\label{eq:vlasovs} 
{\left[ F_0^{(k)}, {\cal K}_k \right]} 
\equiv 0, 
\end{equation} 
\begin{equation}\label{eq:hamil} 
{\frac {\partial S_k} {\partial \theta}} + 
\epsilon H_k {\left( q, {\frac {\partial S_k} 
{\partial q}}; \theta \right)} = 
{\cal K}_k {\left( q, P \right)}, 
\end{equation} 
\begin{equation}\label{eq:poiss} 
{\frac {\partial^2 V_k} {\partial q^2}} = 
4 \pi \! \! \int \! \! dP 
{\frac {\partial^2 S_k} {\partial q \partial P}} 
F_0^{(3-k)} {\left( q, P \right)}, 
\end{equation} 
\noindent 
where $\epsilon$ is formally a small parameter, which will be set 
equal to unity at the end of the calculation. The next step is to 
expand the quantities $S_k$, ${\cal K}_k$ and $V_k$ in a power series 
in $\epsilon$ as 
\begin{equation}\label{eq:expands} 
S_k = qP + \epsilon G_k^{(1)} + 
\epsilon^2 G_k^{(2)} + \epsilon^3 G_k^{(3)} + 
\dots, 
\end{equation} 
\begin{equation}\label{eq:expandk} 
{\cal K}_k = \epsilon {\cal K}_k^{(1)} + 
\epsilon^2 {\cal K}_k^{(2)} + 
\epsilon^3 {\cal K}_k^{(3)} + \dots, 
\end{equation} 
\begin{equation}\label{eq:expandv} 
V_k = {\widetilde{V}}_k + 
\epsilon V_k^{(1)} + \epsilon^2 V_k^{(2)} + 
\epsilon^3 V_k^{(3)} + \dots, 
\end{equation} 
\noindent 
where 
\begin{equation}\label{eq:poisstilde} 
{\frac {\partial^2 {\widetilde{V}}_k} 
{\partial q^2}} = 4 \pi \! \! \int \! \! dP 
F_0^{(3-k)} {\left( q, P \right)}. 
\end{equation} 
\noindent 
Substitution of the above expansions (\ref{eq:expands}) - 
(\ref{eq:expandv}) into Eqs. (\ref{eq:hamil}) and (\ref{eq:poiss}) 
yields perturbation equations that can be solved successively order 
by order. The results are: 

{\it First Order}: $O(\epsilon)$ 
\begin{equation}\label{eq:hamil1} 
{\cal K}_k^{(1)} {\left( q, P \right)} = 
{\frac {\nu_k} {2}} {\left( P^2 + q^2 \right)} + 
{\frac {\lambda_k} {2 \pi}} 
{\widetilde{V}}_k (q), 
\end{equation} 
\begin{equation}\label{eq:generate1} 
G_k^{(1)} {\left( q, P; \theta \right)} = 
{\frac {i \lambda_k} {2 \pi}} 
{\widetilde{V}}_k (q) 
\sum \limits_{n \neq 0} 
{\frac {e^{i n \theta}} {n}}, 
\end{equation} 
\begin{equation}\label{eq:potential1} 
V_k^{(1)} (q; \theta) \equiv 0. 
\end{equation} 

{\it Second Order}: $O(\epsilon^2)$ 
\begin{equation}\label{eq:hamil2} 
{\cal K}_k^{(2)} {\left( q, P \right)} 
\equiv 0, 
\end{equation} 
\begin{equation}\label{eq:generate2} 
G_k^{(2)} {\left( q, P; \theta \right)} = 
- {\frac {\lambda_k \nu_k} {2 \pi}} P 
{\widetilde{V}}'_k (q) 
\sum \limits_{n \neq 0} 
{\frac {e^{i n \theta}} {n^2}}, 
\end{equation} 
\begin{equation}\label{eq:potential2} 
V_k^{(2)} (q; \theta) = - 
{\frac {\lambda_k \nu_k} {2 \pi}} 
{\widetilde{V}}_k^{(2)} (q) 
\sum \limits_{n \neq 0} 
{\frac {e^{i n \theta}} {n^2}}, 
\end{equation} 
\noindent 
where 
\begin{equation}\label{eq:poisstilde2} 
{\frac {\partial^2 {\widetilde{V}}_k^{(2)}} 
{\partial q^2}} = 
4 \pi {\widetilde{V}}''_k (q) 
\! \! \int \! \! dP 
F_0^{(3-k)} {\left( q, P \right)}. 
\end{equation} 

{\it Third Order}: $O(\epsilon^3)$ In third order we are interested 
in the new Hamiltonian, which is of the form 
\begin{equation}\label{eq:hamil3} 
{\cal K}_k^{(3)} {\left( q, P \right)} = 
{\frac {\lambda_k^2 \nu_k} {4 \pi^2}} 
\zeta (2) {\left[ 
{\widetilde{V}}_k^{\prime 2} (q) - 2 
{\widetilde{V}}_k^{(2)} (q) \right]}, 
\end{equation} 
\noindent 
where $\zeta (z)$ is Riemann's zeta-function 
\begin{equation}\label{eq:zeta} 
\zeta (z) = \sum \limits_{n=1}^{\infty} 
{\frac {1} {n^z}}. 
\end{equation} 

\renewcommand{\theequation}{\thesection.\arabic{equation}}

\setcounter{equation}{0}

\section{THE EQUILIBRIUM DISTRIBUTION FUNCTION}

Since the new Hamiltonian ${\cal K}_k$ is time-independent (by 
construction), the equilibrium distribution function $F_0^{(k)}$ 
[see Eq. (\ref{eq:vlasovs} )] is a function of the new Hamiltonian 
\begin{equation}\label{eq:equilibri} 
F_0^{(k)} (q, P) = {\cal G}_k 
{\left( {\cal K}_k \right)}, 
\end{equation} 
\noindent 
where 
\begin{eqnarray}\nonumber 
{\cal K}_k (q, P) = 
{\frac {\nu_k} {2}} {\left( P^2 \! + 
q^2 \right)} + 
{\frac {\lambda_k} {2 \pi}} 
{\widetilde{V}}_k (q) 
\end{eqnarray} 
\begin{equation}\label{eq:equiham} 
+ {\frac {\lambda_k^2 \nu_k} {4 \pi^2}} 
\zeta (2) {\left[ 
{\widetilde{V}}_k^{\prime 2} (q) - 
2 {\widetilde{V}}_k^{(2)} (q) \right]}. 
\end{equation} 
\noindent 
Integrating Eq. (\ref{eq:equilibri}) over $P$ we obtain a nonlinear 
integral equation of Haissinski type \cite{haissinski} for the 
equilibrium beam density profile $\varrho_0^{(k)}$ 
\begin{equation}\label{eq:density} 
\varrho_0^{(k)} (q) = \! \int \! \! dP 
{\cal G}_k {\left( {\cal K}_k \right)}, 
\end{equation} 
\noindent 
where 
\begin{eqnarray}\nonumber 
{\cal K}_k (q, P) = 
{\frac {\nu_k} {2}} {\left( P^2 \! + 
q^2 \right)} 
\end{eqnarray} 
\begin{equation}\label{eq:equihami} 
+ \lambda_k \! \! \int \! \! 
dq' {\left| q - q' \right|}
\varrho_0^{(3-k)} {\left( q' \right)} 
+ 2 \lambda_k^2 \nu_k \zeta (2) 
{\cal F}_k (q), 
\end{equation} 
\begin{eqnarray}\nonumber 
{\cal F}_k (q) = \int dq' dq'' {\cal Z} 
{\left( q-q', q'-q'' \right)} 
\end{eqnarray} 
\begin{equation}\label{eq:calf} 
\times \varrho_0^{(3-k)} {\left( q' \right)} 
\varrho_0^{(3-k)} {\left( q'' \right)},  
\end{equation} 
\begin{equation}\label{eq:imped} 
{\cal Z} (u, v) = {\rm sgn} (u) 
{\rm sgn} (v) - 2 |u| \delta (v). 
\end{equation} 

\section{CONCLUDING REMARKS}

We have developed a systematic canonical perturbation approach 
that removes rapidly oscillating terms in Hamiltonians of quite 
general form. The essential feature of this approach is the use 
of mixed canonical variables. For this purpose the Vlasov-Poisson 
equations are transformed to mixed canonical variables, and an 
appropriate perturbation scheme is chosen to obtain the equilibrium 
phase space density. It is worthwhile to note that the perturbation 
expansion outlined in the preceding section can be carried out to 
arbitrary order, although higher-order calculations become very 
tedious. 

The canonical perturbation technique has been applied to study 
the one-dimensional beam-beam interaction. In particular, rapidly 
oscillating terms due to the periodic beam-beam kicks have been 
averaged away, yielding a time-independent new Hamiltonian. 
Furthermore, the equilibrium distribution functions have been 
obtained as a general function of the new Hamiltonian, and a coupled 
set of integral equations for the beam densities has been derived. 

\section{ACKNOWLEDGMENTS}

This research was supported by the U.S. Department of Energy.

\end{document}